\documentclass[conference]{IEEEtran}
\usepackage{url}
\usepackage{graphicx}
\usepackage{balance}
\usepackage{subfigure}
\usepackage{color}
\usepackage{times,amsmath,epsfig}
\usepackage{amssymb}
\usepackage{textcomp}
\usepackage[linesnumbered,algoruled,boxed,lined,noend]{algorithm2e}
\usepackage{cite}
\usepackage{todonotes} 

\newcommand{\y}[1]{{\color{black} #1}\normalcolor}

\newcommand{\remove}[1]{}

\newcommand{\ignore}[1]{}

\usepackage{booktabs} 

\begin{document}
	
\title{Data Series Indexing Gone Parallel}
%



\author{\IEEEauthorblockN{Botao Peng (Expected graduation date: August 2020; supervised by: Panagiota Fatourou, Themis Palpanas)
}
	
	\IEEEauthorblockA{LIPADE, Universit{\'e} de Paris, 
		botao.peng@parisdescartes.fr}

}
\maketitle

\begin{abstract}
Data series similarity search is a core operation for several data series analysis applications across many different domains. 
However, the state-of-the-art techniques fail to deliver the time performance required for interactive exploration, 
or analysis of large data series collections.
In this Ph.D. work, we present the first data series indexing solutions, for both on-disk and in-memory data, that are designed to inherently take advantage of
multi-core architectures, in order to accelerate similarity search processing times.
Our experiments on a variety of synthetic and real data demonstrate that our approaches are up to orders of magnitude
faster than the alternatives.  
More specifically, our on-disk solution can answer exact similarity search queries on 100GB datasets in a few seconds, and our in-memory solution in a few milliseconds, which enables real-time, interactive data exploration on very large data series collections.

\end{abstract}

%




\section{introduction}
\label{sec:intro}
An increasing number of applications across many diverse domains continuously 
produce very large amounts of data series\footnote{A data series, or data sequence, 
is an ordered sequence of real-valued points. If the ordering dimension is time then we talk about time series, 
though, series can be ordered over other measures
(e.g., angle in astronomical radial profiles, mass in mass spectroscopy, 
position in genome sequences, etc.).} (such as in finance, environmental sciences, 
astrophysics, neuroscience, engineering,  
and others~\cite{DBLP:journals/sigmod/Palpanas15,fulfillingtheneed,Palpanas2019}),
which makes them one of the most common types of data. 
When these sequence collections are generated (often times composed of a large number of short series~\cite{Palpanas2019,DBLP:journals/dagstuhl-reports/BagnallCPZ19}), users need to query and analyze them (e.g., detect anomalies~\cite{conf/icde/boniol20, series2graph}). 
This process is heavily dependent on data series similarity search 
(which apart from being a useful query in itself, also lies at the core 
of several machine learning methods, such as, clustering, classification, 
motif and outlier detection, etc.)~\cite{lernaeanhydra}.

The brute-force approach for evaluating similarity search queries 
is by performing a sequential pass over the dataset.
However, as data series collections grow larger, scanning the complete dataset 
becomes a performance bottleneck, taking hours or more to complete~\cite{zoumpatianos2016ads}. 
This is especially problematic in exploratory search scenarios, where 
every next query 
depends on the results of previous queries.
Consequently, we have witnessed an increased interest in developing indexing techniques 
and algorithms 
for similarity 
search~\cite{shieh2008sax,rakthanmanon2012searching,wang2013data,isax2plus,dallachiesa2014top,zoumpatianos2016ads,DBLP:conf/icdm/YagoubiAMP17,DBLP:journals/pvldb/KondylakisDZP18,ulisseicde,ulissevldb,coconutpalm,DBLP:journals/vldb/KondylakisDZP19,dpisaxjournal,lernaeanhydra,lernaeanhydra2,Palpanas2019,levchenko}. 

Nevertheless, the continued increase in the
rate and volume of data series production with collections that
grow to several terabytes~\cite{DBLP:journals/sigmod/Palpanas15} renders existing data series indexing
technologies inadequate. 
For example, the current state-of-the-art
index, ADS+~\cite{zoumpatianos2016ads} requires more than 4min to answer any single exact query on a moderately sized 250GB sequence collection. 
Moreover, index construction time also becomes a significant bottleneck in the analysis process~\cite{lernaeanhydra}. 
Thus, traditional solutions and systems are inefficient at, or incapable of managing and processing the existing voluminous sequence collections. 

\noindent{\bf{[Contributions]}} 
In our work, we focus on the use of multi-core and multi-socket architectures, as well Single Istruction Multiple Data (SIMD) computations, to accelerate data series similarity search. 
Our contributions are organized as follows.

1. ParIS~\cite{peng2018paris} is the first data series index designed for
multi-core architectures. We describe parallel algorithms for index creation and
exact query answering, which employ parallelism in reading the
data from disk and processing them in the CPU. 

2.ParIS+~\cite{parisplus} is an improvement of ParIS that achieves perfect overlap between the disk I/O and the CPU costs (completely masking out CPU cost) when creating the index.

3. MESSI~\cite{peng2020messi} is the first parallel in-memory data series index. 
Contrary to ParIS/ParIS+, MESSI employs a tree-based query answering strategy that minimizes the number of distance calculations, 
leading to highly efficient search.

\section{Preliminaries}

\begin{figure}[tb]
	\centering
	\subfigure[\y{a} raw data series\label{fig:saxa}] {
		\hspace{-1em}
		\includegraphics[page=1,width=0.40\columnwidth]{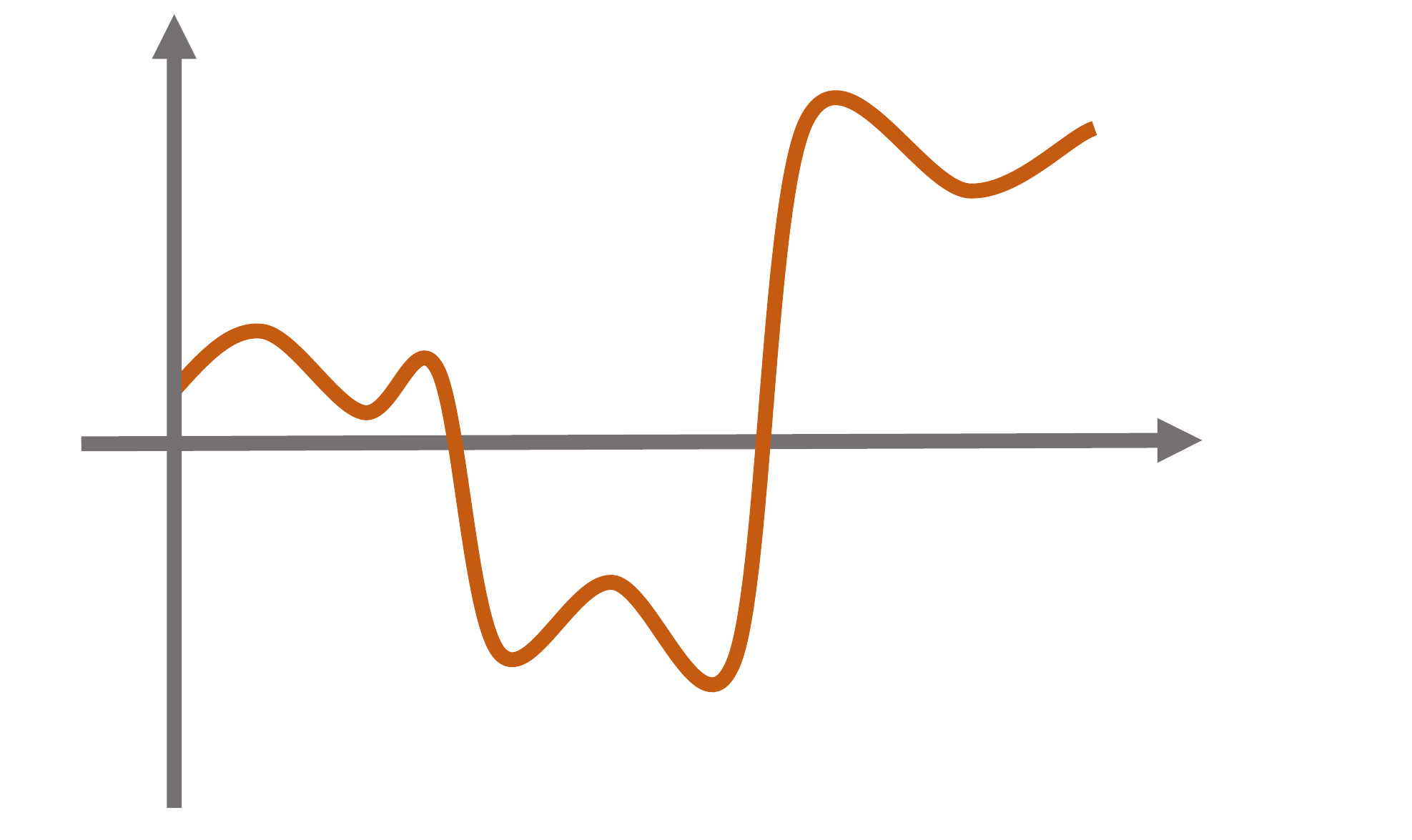}
	}
	\hspace*{0.5cm}
	\subfigure[\y{its} PAA representation\label{fig:saxb}] {
		\hspace{-1em}
		\includegraphics[page=2,width=0.40\columnwidth]{picture/sax2}
	}\\
\hspace*{-0.1cm}
	\subfigure[\y{its} iSAX representation\label{fig:saxc}] {
		\hspace{-1em}
		\includegraphics[page=3,width=0.40\columnwidth]{picture/sax2}
	}
	\hspace*{0.3cm}
	\subfigure[ADS+ index\label{fig:ads}] {
		\hspace{-1em}
		\includegraphics[width=0.45\columnwidth]{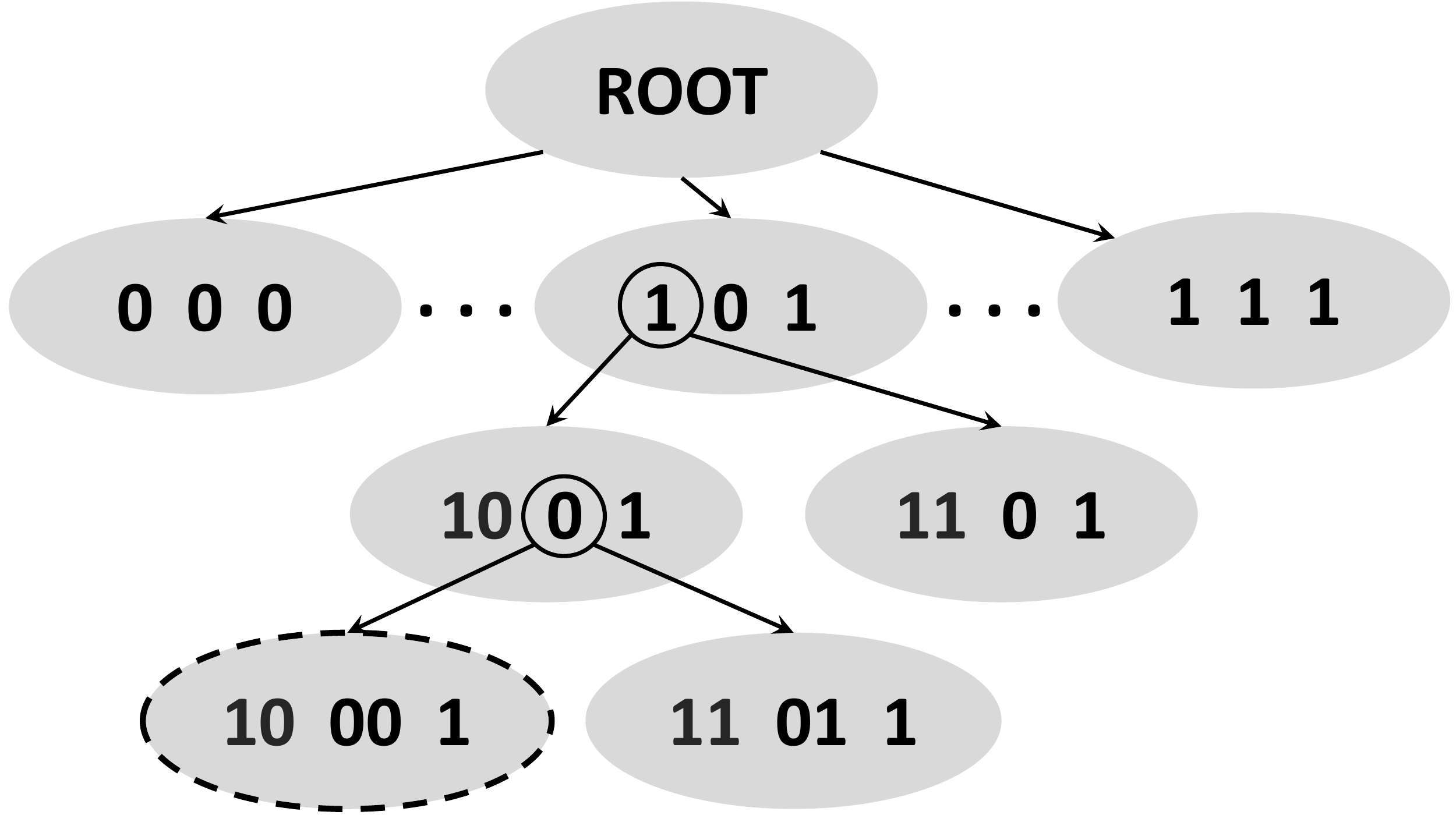}
	}
	\caption{The iSAX representation, and the ADS+ index structure}
	\vspace*{-0.4cm}
\end{figure}

\noindent{\bf [Data Series]}
A data series, $S=\{p_1, ..., p_n\}$, is 
a sequence of points (Figure~\ref{fig:saxa}), 
where each point $p_i=(v_i,t_i)$, $1 \le i \le n$,
is a real value $v_i$ at a position $t_i$ that corresponds to the order of this value in the sequence. 
We call $n$ the \emph{size}, or \emph{length} of the series.
(For streaming series, we create and index subsequences of length $n$ using a sliding window.)

\noindent{\bf [Similarity Search]}
Nearest Neighbor (NN) queries are defined as follows: given a query series $S_q$ of length $n$,  
and a data series collection $\mathcal{S}$ of sequences of the same length, $n$, 
we want to identify the series $S_c \in \mathcal{S}$ 
that has the smallest distance to $S_q$ among all the series in the collection $\mathcal{S}$.
Common distance measures for comparing data series are Euclidean Distance (ED)~\cite{Agrawal1993} 
and dynamic time warping (DTW)~\cite{rakthanmanon2012searching}.


\noindent{\bf [iSAX Representation]}
The iSAX representation first summarizes the points in the data series using {\em segments} of equal length, where the value of each segment is the mean of the corresponding points (Piecewise Aggregate Approximation (PAA)), and then divides the (y-axis) space in different regions by assigning a bit-wise symbol to each region, and represents each segment of the series with the symbol of the region the segment falls into.
This forms a word like $10_2 00_2 11_2$ (subscripts denote the number of bits used to represent the symbol of each segment), which is called the indexable Symbolic Aggregate approXimation (iSAX)~\cite{shieh2008sax}. 



\noindent{\bf [ADS+ Index]}
Based on this representation, the ADS+ index was developed~\cite{zoumpatianos2016ads}. 
It makes use of variable cardinalities (i.e., variable degrees of precision for the symbol of each segment; in order to build a hierarchical tree index, consisting of three types of nodes:
(i) the root node points to several children nodes, $2^w$ in the worst case (when the series in the collection cover all possible iSAX representations), where $w$ is the number of segments;
(ii) each inner node contains the iSAX representation of all the series below it, and has two children; and 
(iii) each leaf node contains the iSAX representations and pointers to the raw data for all series inside it. 
When the number of series in a leaf node becomes greater than the maximum leaf capacity, the leaf splits: it becomes an inner node and creates two new leaves, by increasing the cardinality of the iSAX representation of one of the segments (the one that will result in the most balanced split of the contents of the node to its two new children~\cite{isax2plus,zoumpatianos2016ads}). 
The two refined iSAX representations (new bits set to \textit{0} and \textit{1}) are assigned to the two new leaves. 
In our example, the series of will be placed in the outlined node of the index.

The ParIS/ParIS+ and MESSI indices use the iSAX representation and basic ADS+ index structure~\cite{evolutionofanindex}, but implement algorithms specifically designed for multi-core architectures.

\section{Proposed Solution}
\label{sec:indexcreation}

\noindent{\bf [ParIS/ParIS+ Approach]}
We describe our approach, called Parallel Indexing of Sequences (ParIS), 
and then present ParIS+, which improves uppon ParIS. 

\begin{figure*}[tb]
	\centering
	\includegraphics[page=1,width=0.95\textwidth]{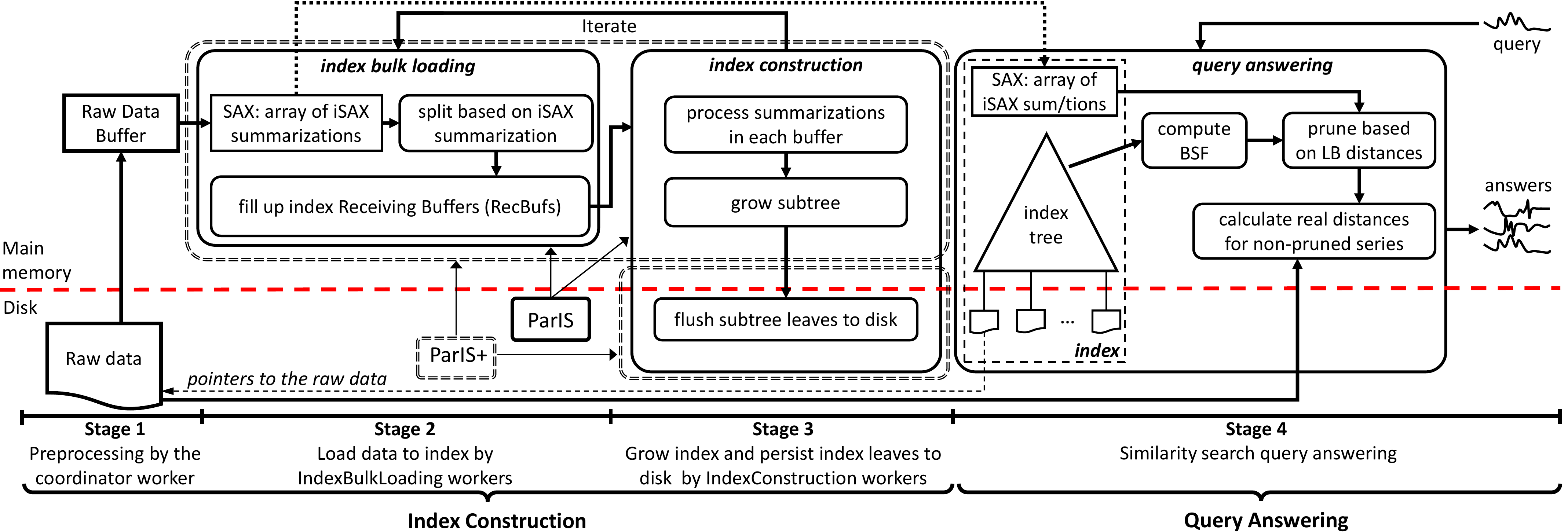}
	\caption{Pipeline for creation of and query answering with the ParIS/ParIS+ indices.}
	\label{fig:workflow}
\end{figure*}

Figure~\ref{fig:workflow} provides an overview
of the entire pipeline of how the ParIS index is created and then used for query answering. 
In Stage~$1$ of the pipeline,
a thread, called the {\em Coordinator} worker, reads raw data series from the disk 
and transfers them into the {\em raw data buffer} in main memory.
In Stage~$2$,
a number of {\em IndexBulkLoading} workers, process the data series in the raw data buffer
to create their iSAX summarizations.  
Each iSAX summarization determines to which root subtree of the tree index the series belongs.
Specifically, this is determined by the first bit of each of the $w$ segments of the iSAX summarization. 
The summarizations are then stored in one of the index Receiving Buffers (RecBufs) in main memory. 
There are as many RecBufs as the root subtrees of the index tree, each one storing
the iSAX summarizations that belong to a single subtree. This number is usually a few tens of thousands (and at most $2^w$, where $w$ is the number of segments in the iSAX representation of each time series; we set $w$ to $16$, as in previous studies~\cite{zoumpatianos2016ads}).
The iSAX summarizations are also stored 
in the array SAX (used during query answering). 

\begin{figure}[tb]
	\centering
	\includegraphics[page=1,width=0.45\textwidth]{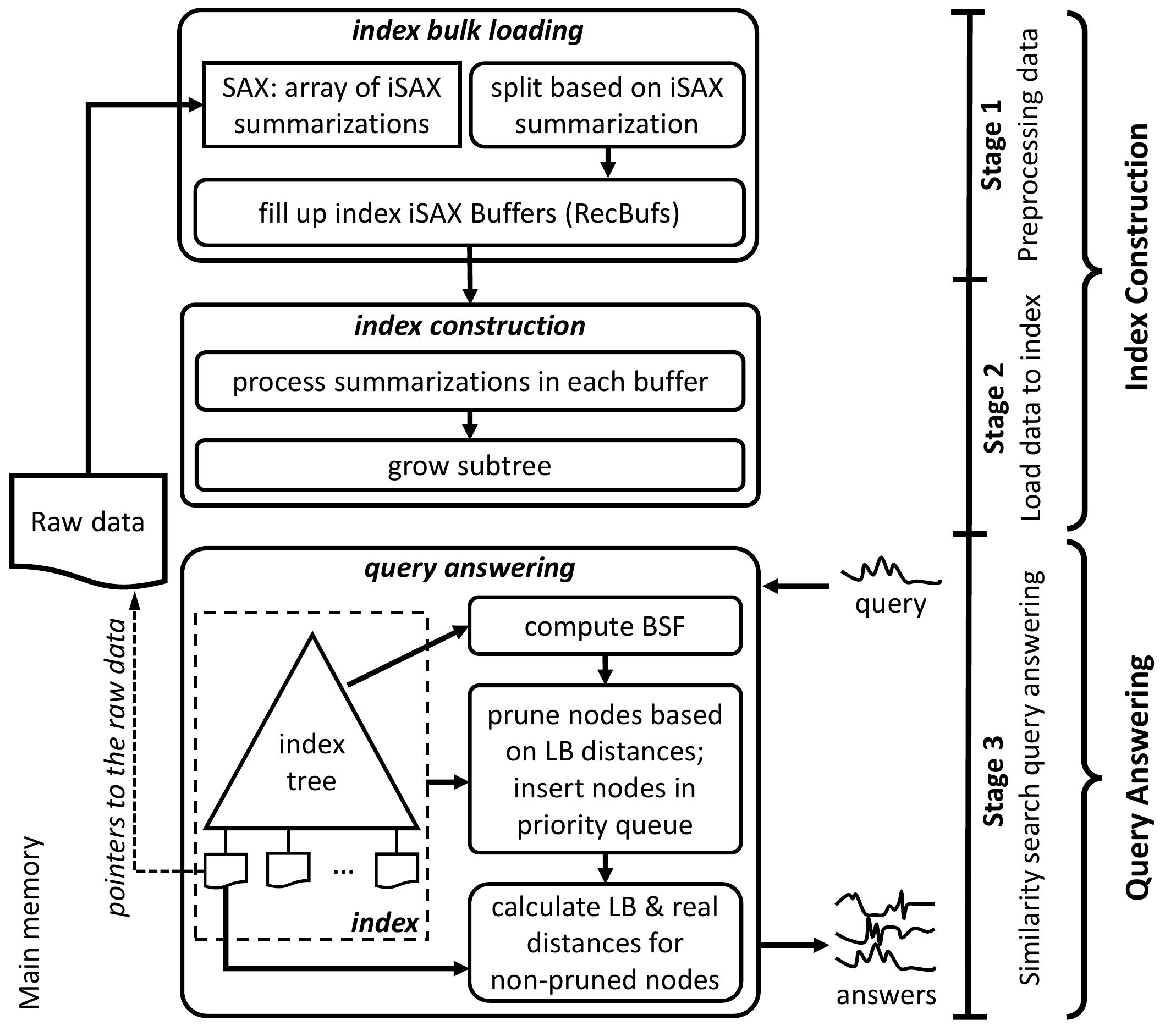}
	\caption{Pipeline for creation of and query answering with the MESSI index.}
	\vspace*{-0.5cm}
	\label{fig:workflow2}
\end{figure}

When all available main memory is full, Stage~$3$ starts. In this stage, 
a pool of {\em IndexConstruction} workers processes the contents of RecBufs;
every such worker is assigned a distinct RecBuf at each time: it reads the data stored in it
and builds the corresponding index subtree. 
So, root subtrees are built in parallel. The leaves of each subtree is flushed to the disk at the end of the 
tree construction process. 
This results in free space in main memory. 
These 3 stages are repeated until all raw data series are read from the disk,
the entire index tree is constructed, and the SAX array is completed. 
The index tree together with SAX form the ParIS index,
which is then used in Stage~$4$ for answering similarity search queries.

Paris+ improves ParIS by completely masking out the CPU cost when creating the index.
This is not true for ParIS, whose index creation (stages 1-3) is
not purely I/O bounded. 
The reason for this is that, in ParIS, the IndexConstruction workers do not
work concurrently with the Coordinator worker. Moreover, the
IndexBulkLoading workers do not have enough CPU work to do
to fully overlap the time needed by the Coordinator worker to read
the raw data file.
ParIS+ is an optimized version of ParIS,
which achieves a complete overlap of the CPU computation with
the I/O cost. In ParIS+, the IndexBulkLoading workers have
undertaken the task of building the index tree, in addition to
performing the tasks of stage 2. The IndexConstruction workers
now simply materialize the leaves by flushing them on disk.

For query answering, ParIS and ParIS+ are the same, and the algorithm operates as follows. 
It first computes an approximate answer by calculating BSF (Best-So-Far), i.e., the real distance between the query and the best candidate series, 
which is in the leaf with the smallest lower bound distance to the query. 
Then, a number of lower bound calculation workers compute the lower bound distances 
between the query and the iSAX summary of each data series in the dataset, which are stored in the \emph{SAX array},
and prune the series whose lower bound distance is larger than the approximate real distance computed earlier.
The data series that are not pruned, are stored in a candidate list, which then a number of real distance calculation workers consume in parallel 
and compute the real distances between the query and the series stored in it
(for which the raw values need to be read from disk).
For details see~\cite{peng2018paris}.

\noindent{\bf{[MESSI Approach]}}
In Figure~\ref{fig:workflow2}, we present the pipeline of the in-MEmory
data SerieS Index (MESSI)~\cite{peng2020messi}. 
In contrast to ParIS/ParIS+,
the raw data are stored in an in-memory array, called $RawData$. 
In Stage 1, this array is split into a predetermined number of chunks. 
A (different) number of {\em index bulk loading worker} threads process the chunks
to calculate the iSAX summaries of the raw data series they store.
Chunks are assigned to index workers one after the other (using Fetch\&Inc).
Based on the iSAX representation, we can figure out 
in which subtree of the index tree an iSAX summary will be stored.
A number of {\em iSAX buffers}, one for each root subtree of the index tree, 
contain the iSAX summaries to be stored in that subtree. 
Each index worker stores the iSAX summaries it computes in the appropriate iSAX buffers. 
To reduce synchronization cost, 
each iSAX buffer is split into parts and each worker works on its own part\footnote{
We also tried an alternative technique: each buffer 
was protected by a lock and many threads were accessing each buffer.
However, this resulted in worse performance due to 
contention in accessing the iSAX buffers. 
}.
The number of iSAX buffers is usually a few tens of thousands and at most $2^w$, 
where $w$ is the number of segments in the iSAX summaries 
of each data series ($w$ is fixed to $16$ in this paper, 
as in previous studies~\cite{zoumpatianos2016ads,peng2018paris}).

MESSI's index construction phase is different from ParIS. 
ParIS uses a number of
buffers to temporarily store pointers to the iSAX summaries of
the raw data series before constructing the  index\cite{peng2018paris,parisplus}. 
MESSI allocates smaller such buffers per thread for storing the iSAX summaries themselves. 
In this way, it eliminates
the synchronization cost in accessing the iSAX buffers. 
To achieve
load balancing, MESSI splits the array storing the raw data series
into small blocks, and assigns blocks to threads in a round-robin
fashion.

When the iSAX summaries for all raw data series have been computed, we move to Stage 2, and the index workers proceed in the construction of the tree index. 
Each worker is assigned an iSAX buffer to work on (using Fetch\&Inc). 
Each worker reads the data stored in (all parts of) 
its assigned buffer and builds the corresponding index subtree. 
Therefore, all index workers process distinct subtrees of the index, 
and can work in parallel and independently from one another, with no need for synchronization\footnote{
Parallelizing the processing inside each one of the index root subtrees would require 
a lot of synchronization due to node splitting. 
}.
When an index worker finishes with the current iSAX buffer it works on, 
it continues with the next iSAX buffer that has not yet been processed.

When the series in all iSAX buffers have been processed, 
the tree index has been built and can be used to answer similarity search queries, as shown in Stage~3.
To answer a query, we first perform a search for the query iSAX summary in the tree index.
This returns a leaf whose iSAX summary has the closest distance
to the iSAX summary of the query.
We calculate the real distances of the (raw) data series pointed to by the elements
of this leaf to the query series, and store the minimum distance in a shared BSF (Best-So-Far) variable. 
Then, the index workers traverse the index subtrees
(one after the other) using BSF to decide 
which subtrees will be pruned. 
The leaves of the subtrees that cannot be pruned are placed (along with the lower bound distance between the raw values of the query series and the iSAX summary of the leaf node) into a number of minimum priority queues. 
Each thread inserts elements in the priority queues in a round-robin fashion
so that load balancing is achieved. 
the same number of elements). 

As soon as the necessary elements have been placed in the priority queues,
each index worker chooses a priority queue to work on, and repeatedly pops leaf nodes, on which it performs the following operations. 
It first checks whether the lower bound distance stored in the priority queue is larger than the current BSF: if it is then we are certain that the leaf node does not contain possible answers and we can prune it; otherwise, the worker needs to examine the series in this leaf node, by first computing lower bound distances using the iSAX summaries, and if needed also the real distances using the raw values. 
During this process, we may find a series with a smaller distance to the query, in which case we  update the BSF.
When a worker reaches a node whose distance
is bigger than the BSF, it gives up this priority queue
and starts working on another, because it is certain
that all the other elements in the abandoned queue have an even higher distance to the query series. 
This process is repeated until
all priority queues are processed. 
At the end of the calculation,
the value of BSF is returned as the query answer. 

Note that, similarly to ParIS/ParIS+, MESSI uses SIMD (Single-Instruction Multiple-Data) 
for calculating the distances of the index iSAX summaries 
from the query iSAX summary ({\em lower bound distance calculations}), 
and the raw data series from the query data series ({\em real distance calculations})~\cite{peng2018paris}.

\section{Experimental Evaluation}
\label{sec:experiments}

We summarize the performance results for index creation and query answering using the ParIS/ParIS+ and MESSI indices, for both on-disk and in-memory data.
We compare our methods to the state-of-the-art index, ADS+~\cite{zoumpatianos2016ads}, and serial scan method, UCR Suite~\cite{rakthanmanon2012searching}. 
We use \emph{two} sockets and split the number of cores equally between them.
The datasets we use are Synthetic (random walk: 100M series of 256 points), SALD (electroencephalography: 200M series of 128 points), and Seismic (seismic activity: 100M series of 256 points).
 

\begin{figure*}[tb]
	\centering
	\begin{minipage}[b]{0.62\columnwidth}
		\includegraphics[page=1,width=\columnwidth]{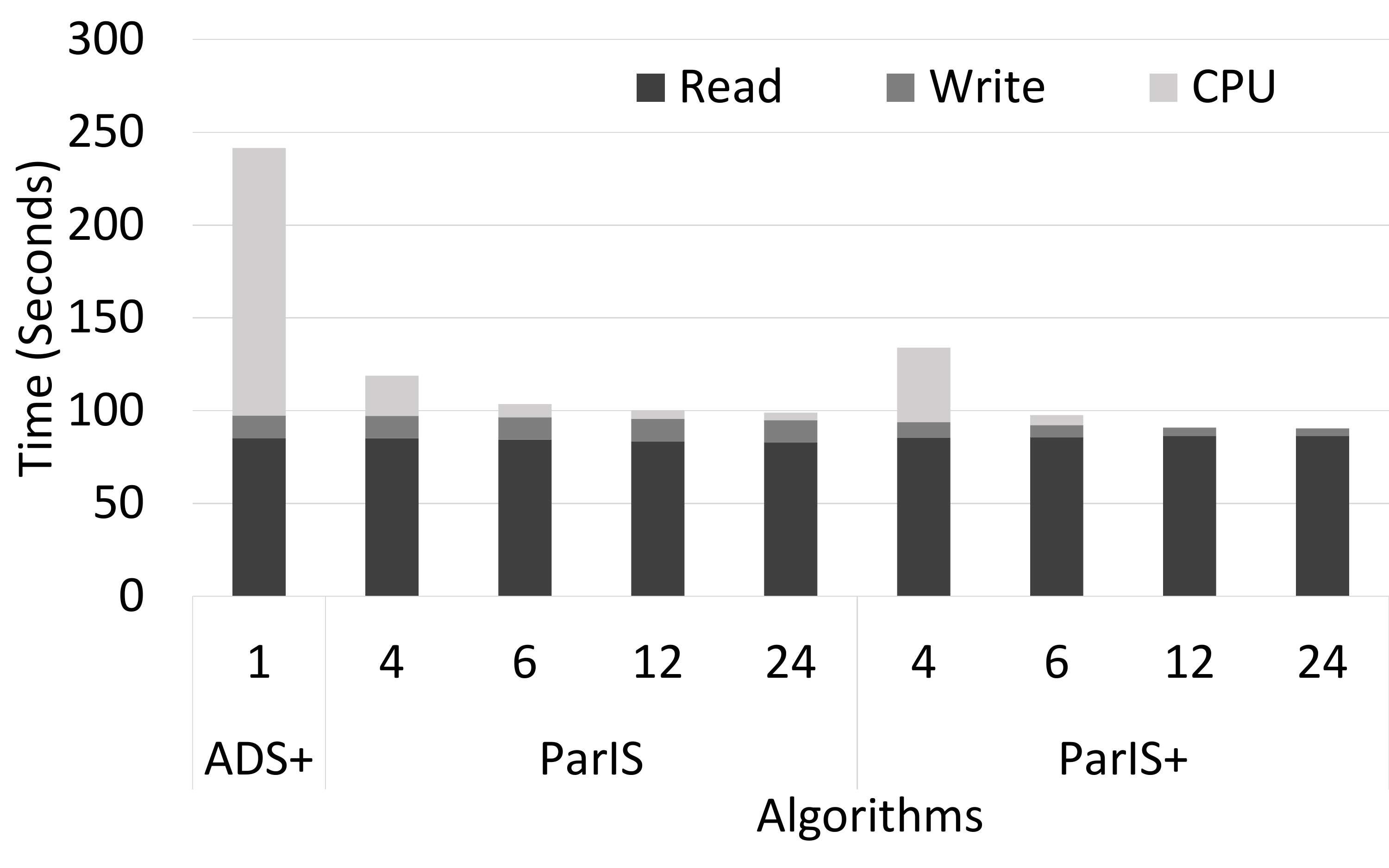}
		\caption{ParIS/ParIS+ index creation}
		\label{fig:vincParIS}
	\end{minipage}
\hspace*{0.1cm}
	\begin{minipage}[b]{0.38\columnwidth}
		\includegraphics[page=1,width=0.8\columnwidth]{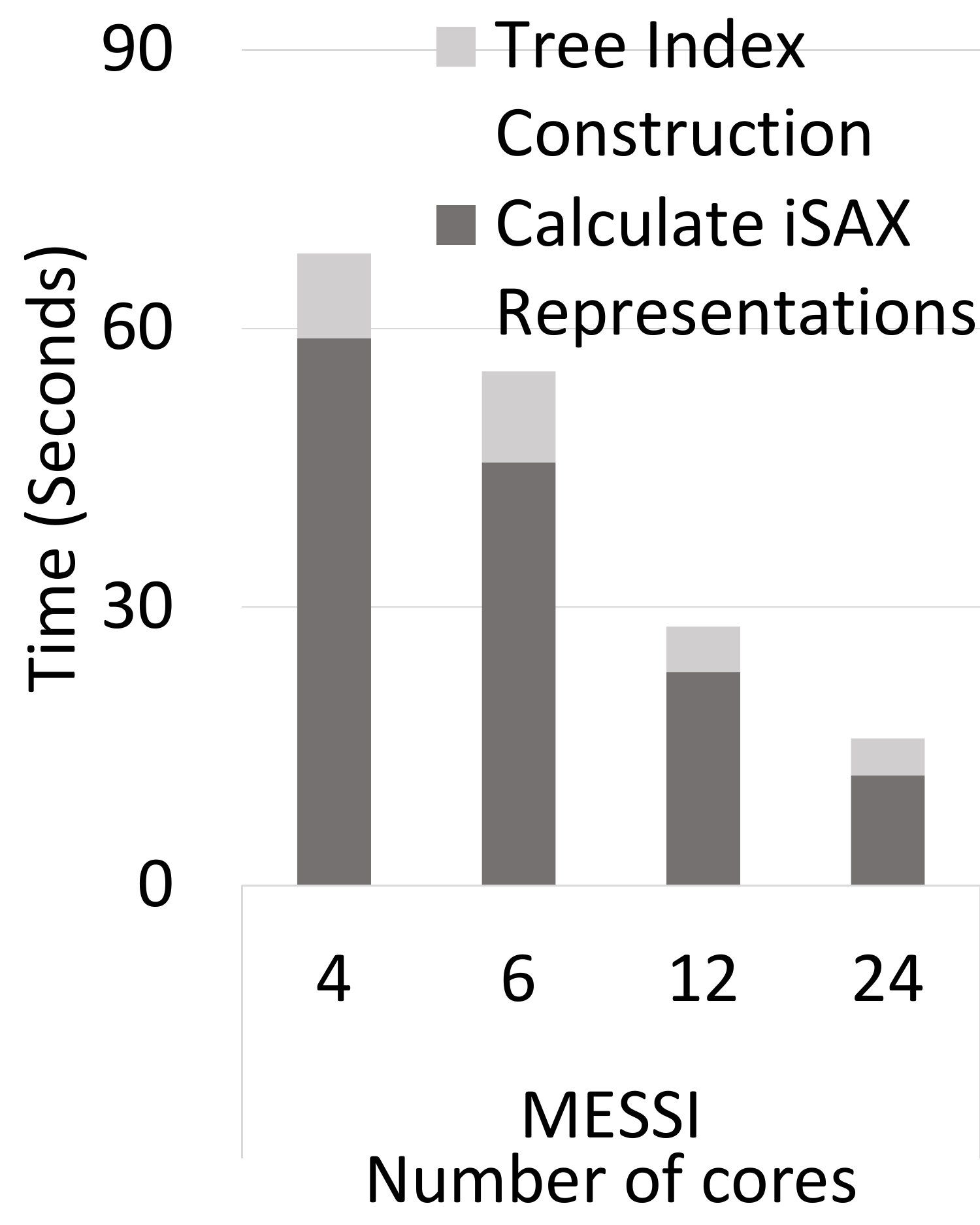}
		\caption{MESSI index creation}
		\label{fig:vincMESSI}
	\end{minipage}
\hspace*{0.1cm}
		\begin{minipage}[b]{0.48\columnwidth}
		\includegraphics[page=1,width=0.95\columnwidth]{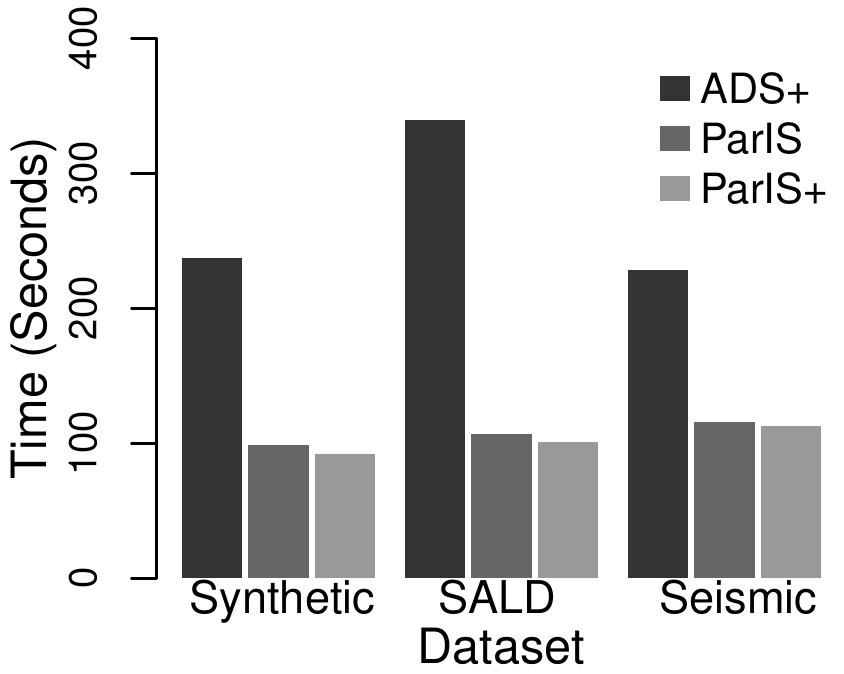}
		\caption{ParIS/ParIS+ index creation}
		\label{fig:incParIS}
	\end{minipage}
\hspace*{0.1cm}
	\begin{minipage}[b]{0.4\columnwidth}
		\includegraphics[page=1,width=0.95\columnwidth]{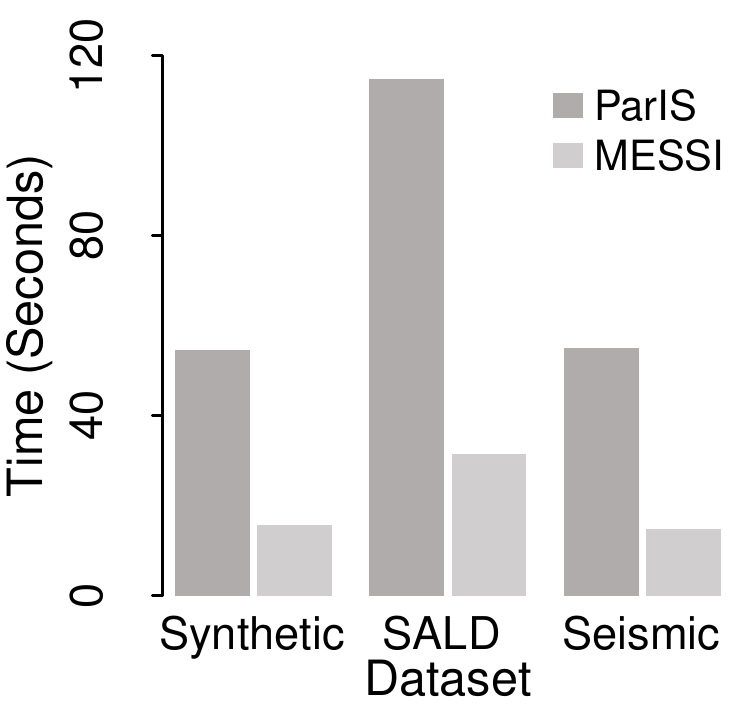}
		\caption{MESSI index creation}
		\label{fig:incMESSI}
	\end{minipage}
\end{figure*}

\noindent{\bf [Index Creation Performance]}
In Figure~\ref{fig:vincParIS}, we evaluate the time it takes
to create the tree index for a Synthetic dataset of 100M series as we vary the number of cores.
The results demonstrate that ParIS+ completely removes the (visible) CPU cost when we use more than 6 cores.
Figure~\ref{fig:vincMESSI} shows that the index creation
time of MESSI reduces linearly as the number of cores
increases (dataset of 100GB). 

We also evaluate the time it takes to create the data series for different datasets of size 100GB.
The results depicted in Figure~\ref{fig:incParIS} show that 
ParIS+ is 2.6x faster than ADS+ for Synthetic, 3.2x faster for SALD, and 2.3x faster for Seismic. 

Figure~\ref{fig:incMESSI} focuses on in-memory index creation. 
We observe that for the 100GB Synthetic
dataset, MESSI performs 3.6x faster than an in-memory implementation of ParIS. 
Note that ParIS is faster than ParIS+ for in-memory index creation (remember that in query answering, they both use the same algorithm and perform equally). 
The reason is that ParIS+ accesses repeatedly the  nodes that are children of the root in order to grow the corresponding sub-trees (for on-disk data, this helps to overlap the CPU with the disk I/O cost). 
However, when in-memory, there is no expensive disk I/O cost to overlap them with; thus, ParIS+ ends up performing unnecessary calculations as it traverses the sub-trees over and over again.
In contrast, ParIS only accesses the children of the root once for every time the main memory gets full (refer to Stage~2 of Figure~\ref{fig:workflow}).
Regarding the real datasets, MESSI is 3.6x faster than ParIS on SALD, and 3.7x faster than ParIS 
on Seismic (both datasets are 100GB in size).

\begin{figure*}[tb]
	\vspace*{-0.60cm}
	\centering
		\begin{minipage}[b]{0.42\columnwidth}
		\includegraphics[page=1,width=\columnwidth]{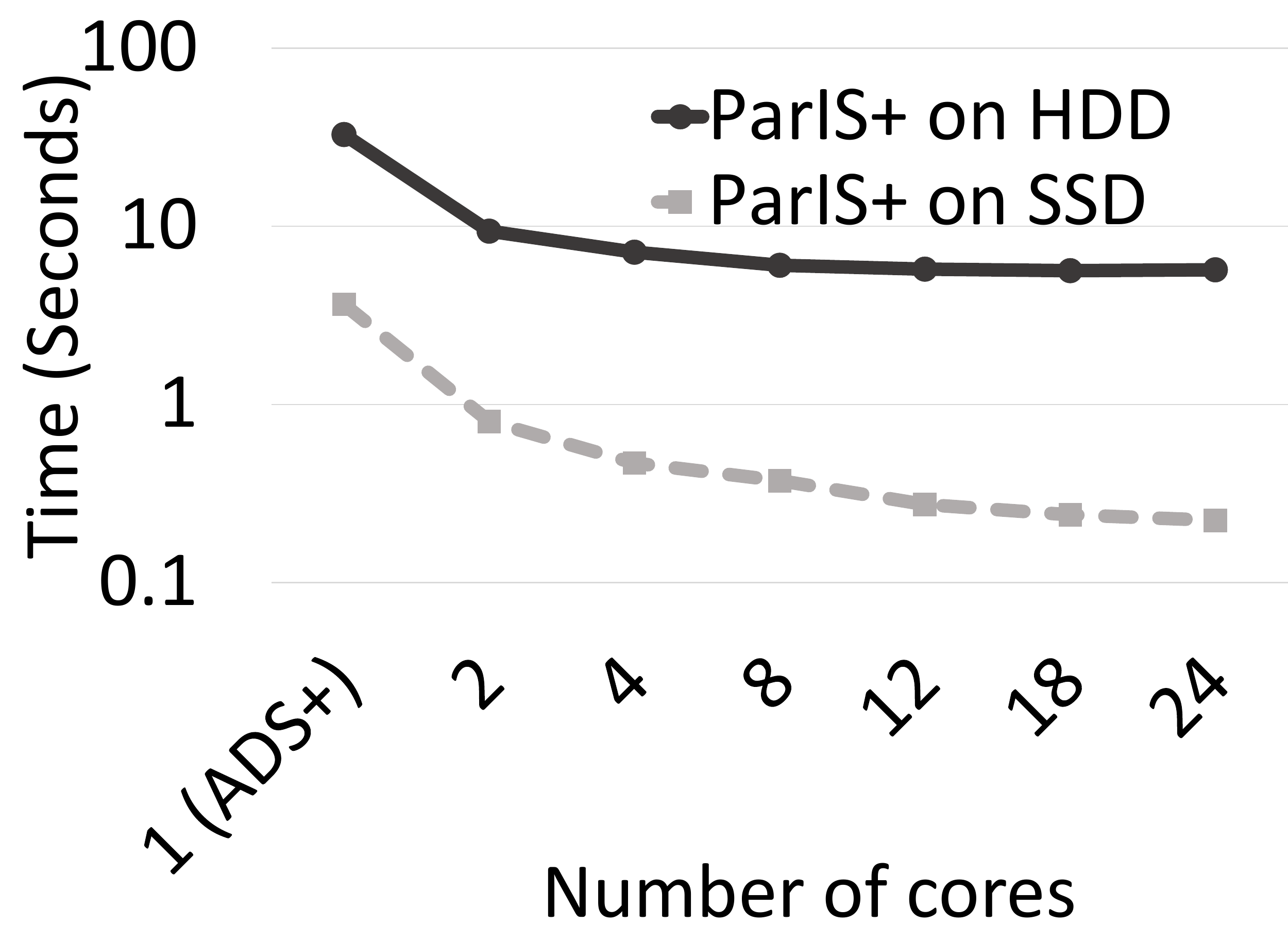}
		\caption{ParIS+ query answering on HDD\&SSD}
		\label{fig:vqparishdd}
	\end{minipage}
	\begin{minipage}[b]{0.42\columnwidth}
		\includegraphics[page=1,width=\columnwidth]{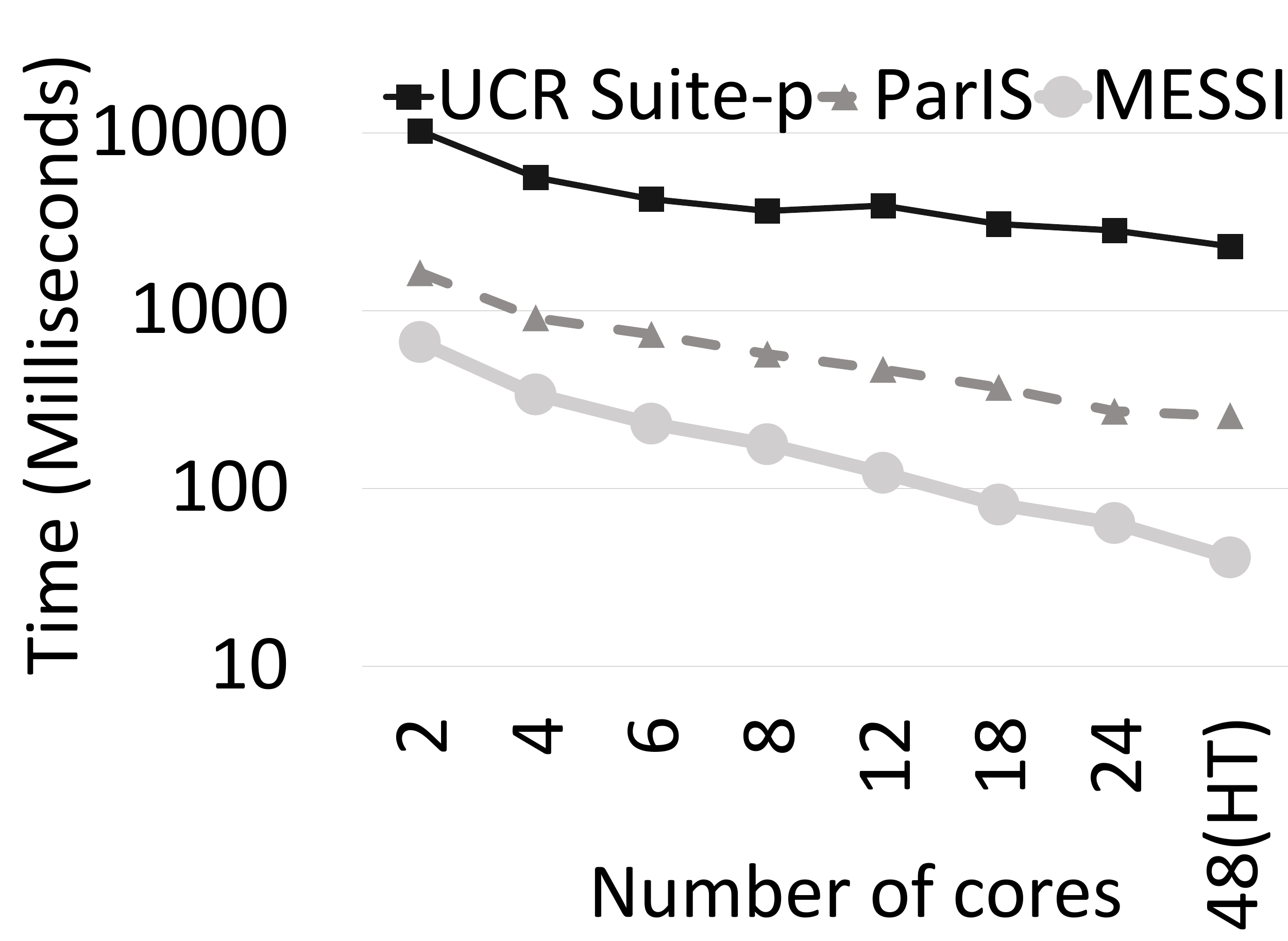}
		\caption{MESSI query answering (in-memory)}
		\label{fig:vqmessi}
	\end{minipage}
	\begin{minipage}[b]{0.38\columnwidth}
		\includegraphics[page=1,width=0.95\columnwidth]{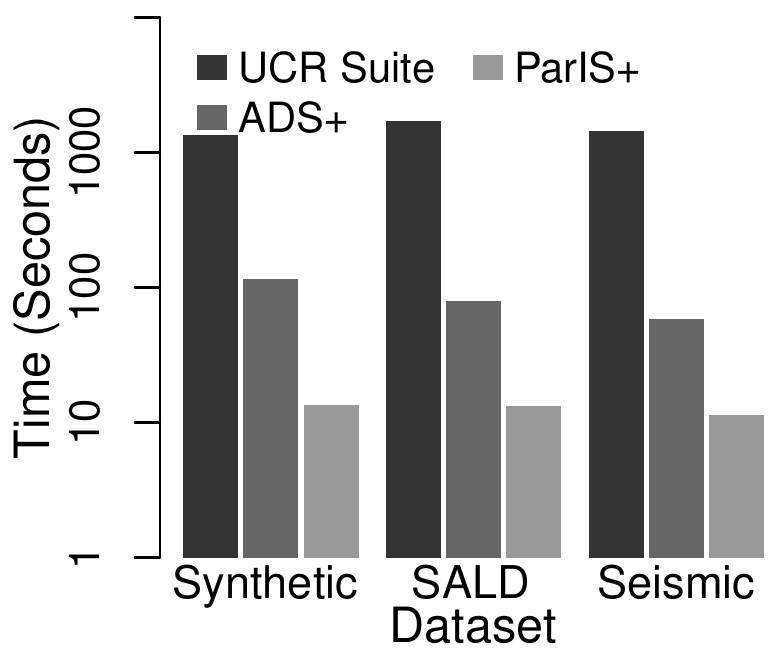}
		\caption{ParIS+ query answering on HDD}
		\label{fig:qparishdd}
	\end{minipage}
	\begin{minipage}[b]{0.38\columnwidth}
		\includegraphics[page=1,width=0.95\columnwidth]{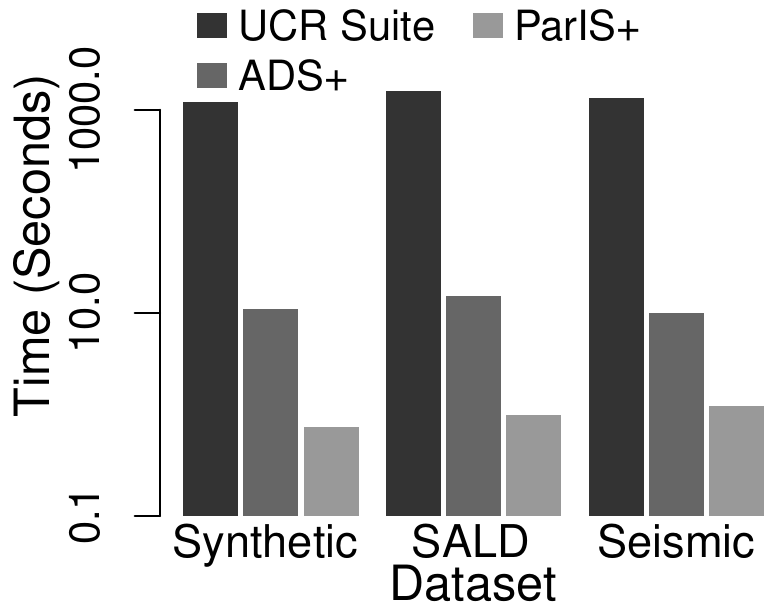}
		\caption{ParIS+ query answering on SSD}
		\label{fig:qparisssd}
	\end{minipage}
	\begin{minipage}[b]{0.38\columnwidth}
		\includegraphics[page=1,width=0.95\columnwidth]{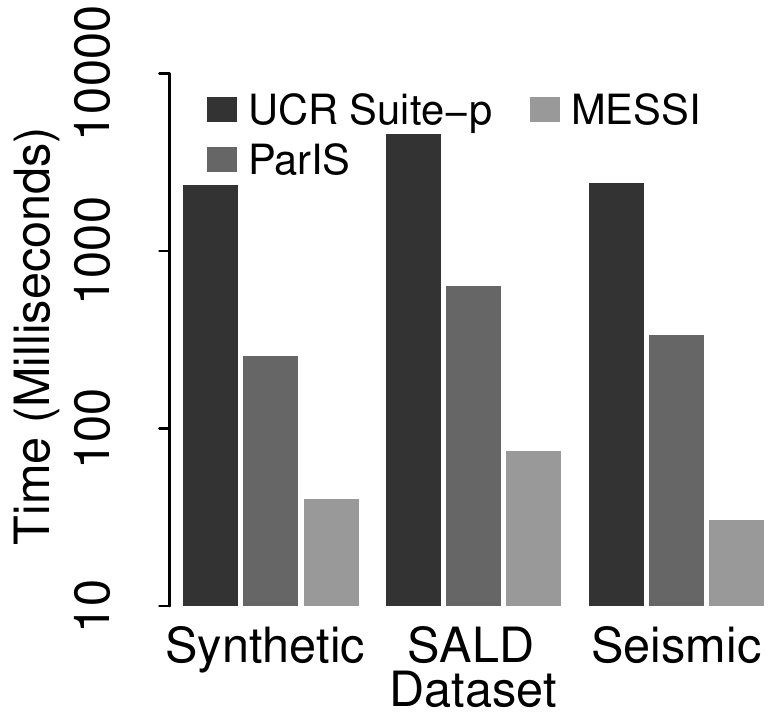}
		\caption{MESSI query answering}
		\label{fig:qmessi}
	\end{minipage}
\end{figure*}

\noindent{\bf [Query Answering Performance]}
Figure~\ref{fig:vqparishdd} (log-scale y-axis) shows the exact query answering time for ParIS+ and ADS+, as we vary the number of cores, for HDD and SSD storage. 
In both cases performance improves as we increase the number
of cores, with the SSD being $>1$ order of magnitude faster.

Figure~\ref{fig:vqmessi} (log-scale y-axis) compares the performance
of the MESSI query answering algorithm to its competitors,
as the number of cores increases, for a Synthetic dataset
of 100GB. 
The results show that MESSI significantly outperforms ParIS and (an in-memory, parallel implementation of) UCR Suite.

In Figure~\ref{fig:qparishdd} (log-scale y-axis), we show that on HDD and across the datasets used in our study, ParIS+ is up to one order of magnitude
faster than ADS+ in query answering, and more than two orders of magnitude faster than UCR Suite.
When the data are stored on an SSD (refer to Figure~\ref{fig:qparisssd}; log-scale y-axis), both ADS+ and ParIS+ benefit from the low SSD random access latency.
In this case, ParIS+ is 15x faster than ADS+, and 2000x faster than UCR Suite.

The results of in-memory query answering, depicted in Figure~\ref{fig:qmessi} (log-scale y-axis), show 
that MESSI perform considerably better than the other approaches. 
MESSI is 55x faster than UCR Suite and 6.4x faster than (the in-memory implemenation of) ParIS. 
The performance improvement with regards to ParIS is because, in contrast to ParIS, MESSI applies pruning when performing the lower bound distance calculations, and therefore needs less computations overall to execute this phase. 
Moreover, the use of the priority queues result in even higher pruning power. 
As a side effect, MESSI also performs less real distance calculations than ParIS. 

Figures~\ref{fig:qmessi} also shows that MESSI exhibits the best performance for the real datasets, 
SALD and Seismic (both 100GB in size), as well. 
The reasons for this are those explained in the previous paragraphs. 
For the SALD dataset, MESSI query answering is 60x faster than UCR Suite and 8.4x faster than ParIS, whereas for the Seismic dataset, 
MESSI is 80x faster than UCR Suite, and almost 11x faster than ParIS.
Note that MESSI exhibits better performance than UCR Suite in the case of real datasets. 
This is so because working on random data results in better pruning than that on real data. 

\section{Conclusions and Current Work}
\label{sec:conclusions}

In this thesis, we describe the first data series indices that exploit the parallelism opportunities of multi-core and multi-socket architectures, for both on-disk and in-memory data.
The evaluation with synthetic and real datasets demonstrates the efficiency of 
our solutions, which are orders of magnitude faster than the state-of-the-art competitors.
This level of performance achieved by our approaches enable for the first time interactive data exploration on very large data series collections.

As part of our current work, we are extending our techniques (i.e., ParIS+ and MESSI) to support the DTW distance measure. 
In order to do this, no changes are required in the index structure: 
we can index a dataset once, and then use this index to answer both
Euclidean and DTW similarity search queries.

Moreover, we are working on a GPU-based solution, where the CPU and GPU collaborate to answer a query: the CPU handles the index tree traversals and real distance calculations (the raw data do not fit in the GPU memory), while the GPU performs the lower bound distance calculations. 
We are also integrating our techniques with a distributed approach~\cite{dpisaxjournal,levchenko}, which is complementary to the ParIS+ and MESSI solutions. 
Finally, we note that our techniques are applicable 
to \emph{high-dimensional vectors in general} (not just sequences)~\cite{lernaeanhydra2}.
Therefore, we will study applications of our techniques in problems related to deep learning embeddings (which are high-dimensional vectors), such as similarity search for images~\cite{JDH17}.

\noindent{\bf [Acks]}
Work supported by Investir l’Avenir, Univ. of Paris IDEX Emergence en Recherche ANR-18-IDEX-000, Chinese Scholarship Council, FMJH PGMO, EDF, Thales and HIPEAC 4.
Part of work performed while P. Fatourou was visiting LIPADE, and while B. Peng was visiting CARV, FORTH ICS.

\bibliographystyle{IEEEtran}
\bibliography{parisinmemory}

\end{document}